# A comparison, analysis, and provision of methods in identifying types of malware and means of malware detection and protection against them.

Sebastian Grochola
Faculty of Science and Technology
Bournemouth University
Bournemouth, United Kingdom
0000-0002-0536-1097

Andrew Milliner
Faculty of Science and Technology
Bournemouth University
Bournemouth, United Kingdom
0000-0003-4836-3006

*Abstract*— In this research paper, our intent is to outline different types of malware, their means of operation, and how they are detected; in order to protect yourself against such attacks. Varied permission, and limited technical resources mean that detecting malware and such attacks becomes more difficult. With the 'normal' user being limited to the UI, their ability to see what happens in the background is virtually limited to none. Many do not have control on how they distribute permission over the data the applications they use controls, or how that data is stored or distributed. They also do not receive any notification as to whether their data is protected against various attacks and if it has not been attacked already. In this paper, we present evidence on what malware is, how malware operates, different types of malware, and the general means of defense.

## I. Introduction

According to a report by OFCOM, there are 42.5 million active adult users of smartphones in the UK, as of 2021 (1). As for malware, Symantec (2), in 2014 alone identified 317 million of new malware software. In the UK alone, that means each smartphone user is outnumbered by malware 7.458 to 1, however the user is vulnerable to all of them. In this paper we will further outline the differences between types of malware; where the focus shifts to means of attack such as Adware, Fileless Malware, Viruses, Worms, Trojans, Bots, Ransomware, Spyware, Mobile Malware, and Rootkits, as identified by Arctic Wolf (3). Each of these methods of attack is different but equally as damaging, only varying in executing the damage long term, or short term.

Previously speaking, prior to the large technological advancements of the early 21st century, malware was simpler, and was much easier to detect. With the amount of digital data increasing, the malware advanced much further.

Often companies suffer from attacks as their means of defense become obsolete, and the companies which develop the simplest apps used by a large number of users daily, often bottleneck their own work due to the inability to provide large updates, which would often likely fix hypothetical 'holes' in their software, temporarily preventing security breaches which often lead to further attacks. This is not to say that an attack is impossible as an attack can be approached from the source of the software, or the user end of it. New malware has also become self-sufficient in the means of attack where it can automatically become dormant in the device or application, run in the background, or mask the attack and even its existence from the user, anti-virus software, firewall, and other means of protection implemented by the 'normal' user or a company. This paper has been organized in such way where following this introduction to the issue, the following sections will further explore the types of malware, their practical comparison, the means of detection, as well as providing a conclusive argument which will provide advice on means of protection and living with the imminent danger that malware poses to daily life, and how to minimize its impact if it does happen to affect yours.

## II. Types of Malware

Being one of the most common forms of attack, malware is potentially one of the most effective and dangerous hacking software as there are many possibilities with it. One of the reasons why malware is so common is because there are 7 different types of malware that each have different functions and can steal data in different ways; the purpose of the malware is up to its author/user to decide. Malware can also appear anywhere due to general attachment and need for the internet and the ability to download material from online straight onto a computer system.

The malware can come from websites but can also be sent to the person directly. In order to trick people into downloading and installing the system, hackers can use a form of malware called a trojan; this attacking method disguises itself as another file or something that may be of interest/ importance to its victim, this is then downloaded by the victim and grants the program access to execute its malicious software. This type of malware often utilizes the form of a digital back door into the theoretical victim's device, and/or files; hence the name Trojan, stemming from the story of the Trojan Horse.

Another form of malware, that shares certain similarities in means of attack with the Trojan malware, is the computer worm. The key similarity to the trojan appears to come from backdoor access into the system, and remote activation of the malware on the infected hardware; the key difference between worm and trojan malware is that worms can spread from one computer system to another using the network that the infected computer is connected to. The network may be a digital or a physical network, and depending on the severity of the malware, the worm can also utilize a network based upon electricity access (power sockets), shared hardware such as servers, or even physical elements of the hardware it's infecting, such as the graphics processor. The worm has the ability to self-replicate files across a network meaning if one person becomes infected and connects to any network, the malware will self-replicate onto devices other unprotected devices on that network. This type of attack often comes from phishing links, and digital mail, where a user opens a link and becomes infected with a 'buried' worm, and by connecting to other networks, that user the proceeds to infect the entire network unaware of the severity of the action. Another characteristic of worm malware is that it often disguises itself as another element of software and buries itself deep within a system; hence the name 'worm'. The malware proves itself difficult to remove as often initial scans with anti-virus software still may find it difficult to find without a full systematic scan.

Another common form of malware is spyware. Once it has been downloaded and activated, the program invades and spreads a metaphorical net with is strictly focusing on collecting data such as passwords, bank account numbers and details, login details, security question answers i.e., any information that can be used to access 'secure' and private information. That information may be remotely sent back to the person who infected the hardware with the spyware or stored as a database internally which requires the person infecting to retrieve it physically from the location of the hardware. This is often the case where the device is offline, or remote. "Spyware is a new type of potentially unwanted programs (PUP) [1] the goal of which is to monitor users' online behaviors without user consent." (4) these types of program are specifically made to invade the privacy of the user and send the data back to the attacker so they may decide what to do with the information. There are hundreds of reports online where people have experienced spyware and recorded some of the different programs that scammers get their victims to install in order to grant themselves access to their victims computer systems, an example of one of these programs is "CoolWebSearch – This program would take advantage of the security vulnerabilities in Internet Explorer to hijack the browser, change the settings, and send browsing data to its author." (5). People post their negative experiences online to attempt to prevent others from falling into the same traps as they did, however this doesn't always work as a preventative method. An important subset of spyware is keylogging; this malware records every letter that is typed on an infected piece of hardware. The keylogger would store any data typed onto a computer, or any device capable of producing digital text, using a normal text file. A keylogger however also has natural uses, often a keylogger is a legitimate program utilized in schools in order to monitor and track the information students may attempt to access throughout their time at a computer.

Among the seven most common types of malware, while all can be used to attack and harass victims in negative ways, some prove to have positive uses. Adware is a type of software that collects information from you to create personalized advertisements on your devices. The counter argument to the use of adware, although very useful in creating personalized shopping lists, film recommendations, music playlists, and advertisements on most internet sites which display advertising, the means of collecting such data has become an issue. Often the data collected may be sold to a larger company which is trying to push their advertising, alternatively the bad forms of Adware such as 'badbiosOS' or the Adobe Flash exploit, may utilize device accessories such as cameras, microphones, and speakers in order to obtain information on user behavior, or worse steal details (6 and 7). In the case of 'badbiosOS', the malware used soundwaves in order to infect other devices, and reboot, and attack the system BIOS which is the internal operating system behind an operating system which keeps all the components running in sync behind the operating system which every device user is familiar with. The flash exploit allowed advertisers to use an older backdoor, that was created by the Adobe Flash add-on prior to its incorporation in most modern browsers and let the person committing the attack use camera and microphone access to gather sensitive information, as well as keywords which can be used to steal from the user, alongside being used advertising, and attracting other means of malware such as 'spam'. Similar tactics are also used by major companies such as Facebook and Amazon that collect data on your recent searches to provide the user personalized ads and suggestions on things the user may be more interested in. This may be similar to phishing, as well as utilizing adware but these companies often find themselves pushing the legal parameters of the law in order to maximize their abilities to advertise.

The fourth out of the seven types of malware are rootkits. Rootkits used to be the term used to describe a collection of tools that enabled administrator level access to a computer or network; the type of malware that shares the same name provides a similar sort of process where the software grants privileged access to a computer system/ network whilst also masking its presences to prevent itself from being detected. "Rootkit attacks are a serious threat to computer systems. Packaged with other malwares such as worms, viruses and spyware, rootkits pose a more potent threat than ever before by allowing malware to evade detection." (6). In that way it shares similarities to the trojan software that disguises itself to prevent being detected.

Ransomware is one form of malware that has its own unique style of attack, unlike other malware that attempt to hide or destroy files the ransomware malware locks and encrypts files on a system and then holds them for ransom, as it says in the name, in order to gain the decrypt key. The reason that this form of malicious software is effective is because it puts its victims in a position where the easiest and cheapest method to solving the issue is to pay the ransom in order to gain access to the file. "A survey conducted in 2016 with 290 organizations from different industries in the United States, Canada, Germany and the United Kingdom found that nearly 50% of them had been victims of a ransomware attack during the previous 12 months (Osterman Research and Inc., 2016). Around 40% of the targets declared having paid the ransom. However, even if the organization pays the ransom, there is no guarantee that it will recover the files." (7). The statistics that are presented in this article, speaks to how common and easy it is for people and companies to fall victim to this form of malware causing their files and data to be held hostage; it also demonstrates the position that the hackers put their victims in with this virus where 40% of the 290 organizations paid the ransom as it was the easiest option however, as the article states, there is no guarantee that once ransom is paid that they will recover their files. Another risk of dealing with ransomware is regardless of data recovery; dealing with the risk of data distribution, especially if the data stolen and/or locked away by ransomware is sensitive data. Such data is not limited to login details, bank account numbers, stretching all the way into medical records, personnel files, and large corporate accounts. Ransomware proves itself in being one of the most effective methods of attack as it leaves the user with no ability to counteract without the access key, which can only be provided by the attacker. An example of a successful attack would be the December 2021 attack on 'Planned Parenthood', where a centralized database of Planned Parenthood was attacked by ransomware, which exposed, and distributed data for 409,759 active patients (8). Such issue not only posed form malware, or in this case, Ransomware, however this attack may be an example of a wider issue where a business which operates with multiple location, use local copies of files, in a number of location, syncing to a centralized location. If a network as such is infected, the automatic syncing would compromise and infect every other device in the network, leading to such breaches as the Planned Parenthood breach identified earlier.

### III. PRACTICAL COMPARISON

Based on the descriptions and capabilities of the malware listed previously, it has become clear that malware is a subject that requires a lot of research in the hopes that the preventative methods that are developed can put a stop to the effectiveness of the malicious software, working towards eliminating the posed threat by malware almost completely. A brief look into the academic research behind malware led to finding a journal that demonstrated and tabled information that showed that malware prevention has become an intercontinental issue, with publications towards the topic of malware appearing from every accessible continent.

| Geographical areas | Publications (%) |
|---|---|
| North America | 34.7 |
| Asia | 30.6 |
| Europe | 26.5 |
| Middle East | 3.7 |
| Australia | 3.3 |
| South America | 1.0 |
| Africa | 1.0 |

(8)

In the journal they look into the development of Malware and the history of the research into malware using the analytic method of biometrics. "Using "malware" as the main keyword, we identified over 4000 articles and scrutinized before being classified into 2158 main related articles." (8). Only by using "malware" as a keyword they were able to find thousands of studies into the subject, this determines that it is a heavily researched area and if nothing else, this stands as an example of the possibilities with this type of software. The issues that have been mentioned previously is that the malware can do serious harm to an individual computer and depending on the information that is stored on the computer, could cause serious damage to a person's life. This does not mean that the uses of malware are strictly limited to malicious usage.

Applicability-range of various approaches.

| Solution | Techniques | Inside Organization | Outside Organization |
|---|---|---|---|
| End-point solutions | Access-controls (Cheng et al., 2017a), Behavioral monitors, Mobile device security (Anon), trusted execution environments (TEEs) (Kim et al., 2017), Encryption (Zhang et al., 2009) | Applicable | Not applicable |
| Network monitoring | Traffic scanners, Packet inspection, cloud monitors (Liu et al., 2018; Priebe et al., 2014), service hardening | Applicable | Not applicable |
| File System Monitoring | File scanners (Gula and Ranum, 2018), activity scanners (Asaf et al., 2012), behavioral monitors (Guevara et al., 2014) | Applicable | Not applicable |
| Watermarking | File watermarking (Papadimitriou and Garcia-Molina, 2010; Govinda and Joseph, 2017) | Applicable | Not applicable |
| Right Management & Access Control | Access-control, Behavioral monitors (Guevara et al., 2014), Mobile device security, | Applicable | Not applicable |
| DocGuard (this work) | Embedded malicious signature | Applicable | Applicable |

(9)

In this table, researchers have compiled different uses for malware that can have a more positive use; each example has also been followed by research done into that specific function to conclude that each example is more than speculation. Furthermore, the practical uses for malware in a legal sense is a morally agreeable subject and is used in the circumstance of trying to protect data and files rather than to cause damage in an aggressive manner.

That being said the negative uses of malware are a lot more common, in 2015 the ISACA released an estimate the number of cyber-attacks on company to increase by substantial figures "according to the Information Systems Audit and Control Association (ISACA[1]) report,[2] 74% of organizations expect to fall prey to a cyber-attack in 2016. Attacks aimed at organizations" (10). The growing number of cyber-attack were a concerning issue for the industry and with the events in the years following like covid led companies to experience more cyber attacks.

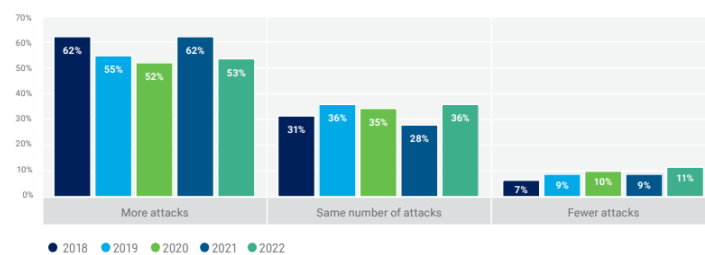

FIGURE 30—YEAR OVER YEAR COMPARISON OF CYBERSECURITY ATTACK REPORTING[15]

(11)

This chart from the ISACA demonstrates that as companies began to return to working after the pandemic, the number of cyberattacks increased by 10% from the previous year, while only 9% of companies reported fewer attacks. With the response to the global pandemic to work online from home, the majority of companies became more vulnerable as they relied on employees to work on external servers that could have potentially been unprotected to cyberattacks. Alongside such danger, working from home also expanded the danger pool to individual malware attacks which may impact not only the sensitive information the company employees are working with but also individual blocks of sensitive data belonging to the user; seeing as very few companies provided company devices for people to avoid working from private hardware. With the many different forms of malware and their largely varied yet extreme extent of capabilities, it is likely that a moderate percentage of these cyberattacks were using some form of malware to assault companies as they resumed business.

In 2021 there were reports of nearly 500 million cyberattacks committed in the first half of 2021 using ransomware software. When considering that ransomware is only 1 of 7 different forms of malware, the number of cyberattacks has the potential to be the biggest crime threat on the planet; with the modern era relying more and more on technology, there is more research and development required in the area of detection and preventative software for both major companies and individual computer systems. An example of one of the biggest malware attacks in 2021 was a ransomware attack committed by the REvil/Sodinokibi gang. In April 2021 an anonymous source posted on a digital crime thread that the gang was about to reveal their "largest attack ever". In the previous month of that year, the same gang had committed a ransomware attack on Acer where they demanded $50 million from the company. In April however, the cyber gang revealed that they had attacked one of the business partners of the tech conglomerate Apple. When that attack failed, the cyber gang turned their attention to Apple by releasing documents they had stolen from Apples business partner. Once again, the gang agreed to a ransom of $50 million on May 1st. (12) Using this as an example it is clear to see that with even major technology companies falling victim to malware attacks, there is a clear demand for more research and development on means of detection in order to prevent both companies and individuals from falling victim to these forms of cybercrime.

## IV. MEANS OF DETECTION

As malware is well established in the cyber security world as a serious issue, that has led to many major companies developing software that is able to detect malicious software and warn a user before the malware gains access to their device. "smart phones normally store private user data such as pictures, messages, and personal credentials. Thus, they become the target of many malicious attackers [30], [15]. In the smart phone industry, devices with Android operating system hold a leading position. However, around 97% of mobile malware target the Android phones. In recent years, security incidents of Android mobile phones occurred frequently, some serious attacks happened also at Apple phones. This situation motivates us to study mobile app security" (13). The issue of malware is an issue that even major technology companies have had their issues with, and has been the subject of serious research in order to determine how to defend against it, not only on computer systems, but on other devices such as mobiles, tablets, and in the modern era, even cars that have computer systems onboard.

Issues with spyware can be identified in a number of different ways "Users infected with spyware commonly experience severely degraded reliability and performance such as increased boot time, sluggish feel, and frequent application crashes." (14). Using software that measures the performance of your computer system is a potential preventative method for releasing sensitive data to an attacker using spyware or keystroke malware. One way to protect against spyware is through applications such as Driverbooster which scans every available driver on a device and ensures that every driver on the device, and updates every available software on the device. This allows the user to keep track of their device's performance and allows them to accurately estimate whether the device is showing any warning signs for spyware such as applications crashing and sluggish clock speed. This combined with antivirus software on a device will reduce the likelihood of the user becoming a victim of a malware attack.

If a hard drive gets infected with malware such as trojan or a rootkit, the hacker could potentially cause damage to the hard drive that would completely disable the device. Equally, the use of a rootkit could leave the victim locked out of their own device. In an instance where too much damage has been caused to the hard drive, to remove the virus, the hard drive may have to be removed and replaced in order to continue using the device. Unfortunately, it would mean that all the data on the hard drive would also be lost, however in most cases where the virus has completely infected the device, it is the only way to continue using the device.

There are companies that have developed antivirus software available for public use, such as McAfee, and Bitedefender. These software packages have been developed to detect and defend computer systems from all known forms or malware and viruses. Most of these forms of antivirus software receive regular updates, in order to stay up to date with defending against known viruses and different types of malware attacks. One of the features is to scan and analyze files that are downloaded onto the computer in order to determine that they are safe to install; this is an effective method as the average user of a computer device would be unaware of what a malicious file would look like and with trojan malware, as long as the file has the right name, they might be very quick to download and install anything that looks correct, while the trojan malware has disguised itself to look like it's the file that the user wants.

One of the key factors in knowing whether a file from a website may or may not contain malware, is knowing whether the website is safe, one of the services that McAfee and other protection services offer is checking websites before you enter them to ensure that they are trusted and safe. In terms of other devices that connect to the internet, apple and android phones also possess a website checker to ensure that there will be no issues with downloading malicious software onto a mobile device or tablet. With companies requiring their data to be available in more locations than ever, a lot of companies are using systems like The Cloud to store their data, this means that there is a completely different way for attackers to target data from big companies; this has therefore forced companies to try and develop new software to detect malware attacks, which has yielded new challenges. "The detection of malware in cloud environments relies on standard security mechanisms and tools, such as anti-virus. The ability of such tools to detect new unknown malware is limited due to their reliance on known malware signatures used for detection." (15). The ability to defend against malware is cut up into 3 stages, detection, prevention and fixing. Detection is the ability to stop malware from being downloaded to a computer system and therefore not risking any potential issues. If that stage has been passed, then the prevention stage is using antivirus software to identify and stop any malware from executing its programming. The final stage is fixing; in this stage, the malware has been executed, and the hardware of the device has to be removed and replaced. In the final stage, the data has been lost and is unsalvageable. Therefore, the best method of preventing malicious use of malware is detection as it prevents the possibility of any known type of malware or virus gaining access to the device.

## V. Conclusion

To conclude, the duration of the paper proved somewhat worthwhile, where we began by focusing on the different types of malware. Identifying the different types of malware with the primary focus being on spyware, adware, ransomware, trojan viruses, and worms, gave an impartial insight to the threat that the user may face. With all data you must have an attempt at protecting it against various threats. The practical comparison within our research had thoroughly proved itself to be vital in identifying the general demographic that suffers from malware attacks, alongside the provision of numbers which provide a perspective on the realities of suffering from a multitude of malware attacks, such as ransomware, and spyware attacks. The overwhelming establishment of malware software is an ongoing battle, with a correlation to the number of updates, and fixes in order to protect against rising threats. Means of detection is an interesting term as although malware may be detected, and be dormant, not actively posing a threat, it could well turn out to be one task detecting it, and another to actively attempt to remove the malware from the device. Our research also showed the sheer size of the market for commercial antivirus software as well as technical provisions created in order to avoid malware attacks specifically on mobile devices. Disabling developer kits, blocking the installation of files form third party sources are all means of detection and defense against malware. Our research presented that although all these means of detection may ascertain a certain level of safety for the user, however it is also safe to infer that all are only temporary means due to the evolutionary nature of malware, and the availability of exploitation due to outdated files, or inferior means of defending.

Such research however only provided us with a definitive threat, without a definitive means of defending against it. The results only provide a temporary measure of defending against malware on a small scale without any further ability to expand. Further work would allow a deeper understanding of malware as well as divert some attention into the psychology of malware attacks; understanding the causality would benefit greatly in pre-determine the effects of the attack, alongside inferring the potential outfall caused by such attack. Another area that would further develop in future work would be the expansion of the practical comparison. By firstly expanding the research on the technical structures of various means of malware, it creates a space for deeper analysis which would break down the skeletal structure of the malware, allowing exposure for the malware to defensive threats, and possible elimination of said threat. Application of this onto our research into the real-world examples of malware attacks creates a space to fill a void which could include a defensive action plan to protect the general masses against malware, alone from raising awareness but creating provisions to tackle such attack if there was one to occur.